\documentstyle[12pt]{article}

\newcommand{\beq}{\begin{equation}}
\newcommand{\eeq}{\end{equation}}
\newcommand{\bea}{\begin{eqnarray}}
\newcommand{\eea}{\end{eqnarray}}
\newcommand{\st}{\theta}
\renewcommand{\ni}{\noindent}

\newcommand{\AmS}{{\protect\the\textfont2
  A\kern-.1667em\lower.5ex\hbox{M}\kern-.125emS}}
\def\lsim{\raise0.3ex\hbox{$<$\kern-0.75em\raise-1.1ex\hbox{$\sim$}}}
\def\gsim{\raise0.3ex\hbox{$>$\kern-0.75em\raise-1.1ex\hbox{$\sim$}}}

\begin{document}
\hfill{\vbox{\hbox{HLRZ 27/93}
             \hbox{LU TP 93-6}
             \hbox{April 1993}\relax}
\vspace{12pt}
\begin{center}
{\large\bf Finite-size Scaling at Phase Coexistence}
\vspace{24pt}

{\sl Sourendu Gupta}

\vspace{12pt}
HLRZ, c/o KFA J\"ulich \\
D-5170 J\"ulich, Germany\\
\vspace{12pt}

{\sl A.~Irb\"ack} and {\sl M.~Ohlsson}

\vspace{12pt}
Department of Theoretical Physics, University of Lund\\
S\"olvegatan 14A, S-223 62 Lund, Sweden\\
\vfill
{\bf Abstract}
\end{center}
\vspace{12pt}
\ni
{}From a finite-size scaling (FSS) theory of cumulants
of the order parameter at phase coexistence points,
we reconstruct the scaling of the moments. Assuming
that the cumulants allow a reconstruction of the free
energy density no better than as an asymptotic expansion,
we find that FSS for moments of low order is still complete.
We suggest ways of using this theory for the analysis of
numerical simulations. We test these methods numerically
through the scaling of cumulants and moments of
the magnetization in the low-temperature phase of the
two-dimensional Ising model.
\vfill
\newpage

\section{INTRODUCTION}

For systems close to a second order
phase transition, finite-size scaling (FSS)
is routinely used to extract thermodynamical
information from systems of fairly small size. An equivalent
theory at phase coexistence points, {\sl i.e.\/}, first order
phase transitions, is clearly of interest. In the past,
qualitative methods were used to identify
and characterise first order transitions.
With improved computer resources these are
inadequate. An useful theory of finite-size scaling
should allow us to extract the couplings at which
the transition occurs, as well as other
dimensional quantities like the latent heat
(or spontaneous magnetisation) and the specific heat
(or the magnetic susceptibility). In principle, the precision
of numbers extracted by such scaling
should then only depend on the computer resources available.

{}From fairly general arguments about the nature of discontinuities
at a first-order phase transition, Fisher and Berker \cite{fb}
obtained the infinite volume limit approached by measurements
performed at finite volumes.
Correction terms were later computed in a phenomenological model
\cite{bl,clb} called the double-Gaussian model. Here the peaks
in the probability distribution for the coexisting phases were
approximated by Gaussians. This model correctly predicts the
first term in a series of corrections
in inverse powers of the volume, $V$, about the leading term
of \cite{fb}. Numerical studies of the two dimensional Ising
model were claimed to be in good agreement with these predictions
\cite{bl}.

The major new developments are due to
Borgs, Koteck\'y and Miracle-Sol\'e \cite{bk,bkm}.
Powerful rigorous results were obtained for the Ising
model at large $\beta$ and $q$-state
Potts models with large $q$.
They provide, for instance, an estimator of
the transition coupling that should have exponentially
small finite-size corrections.
The results are claimed to apply more generally
than the cases they have been proven for.
For shifts in the transition coupling,
numerical studies of Potts models support this
claim \cite{bj,blm}.
However, the agreement with data
for some other quantities \cite{blm,lk} is less convincing.

The basic idea is to decompose the partition function into
a sum of parts, each due to one of the coexisting phases,
and neglect contributions due to phase mixtures.
Each of these parts of the partition function then yield
quantities related to free energies in the pure phase.
The analysis proceeds by expanding these in a power
series around the phase transition point, yielding
expansions of the moments in inverse powers of the volume.
In refs.~\cite{bk,bkm}, the
leading and first correction terms
in such expansions were obtained,
including rigorous bounds on the approximations made.
The leading term is the same as that obtained in ref.~\cite{fb}
and the first correction term agrees with that in
refs.~\cite{bl,clb}. In addition, the results of
refs.~\cite{bk,bkm} resolve an ambiguity concerning the
relative normalisation of the coexisting phases.

Detailed comparisons between numerical data and
the leading and first correction terms in the predictions
have been performed for the temperature-driven
phase transition in the two-dimensional
$q$-state Potts model for $q=7,8$ and 10 \cite{blm,lk}.
As mentioned before, the
agreement was not perfect. A possible reason is that
the lattices were not large enough for the analysis
to be applicable.

This interpretation is substantiated
by recent results for the model with $q=20$ \cite{bnb}.
These authors used the multicanonical algorithm \cite{bn}
to overcome the exponential slowing down associated with
the tunneling between the coexisting phases. As a result,
they were able to study larger system sizes $L/\xi$
($\xi$ is the correlation length) than previously done.
The results show good agreement with
first-order FSS predictions on large lattices.

In order to control finite-size effects,
we start by analysing the structure
of the FSS expansion in powers of $1/V$.
One interesting result is that the volume at which
the leading order in $1/V$ describes the scaling of
data well, depends on the variable being studied.
For the $k$-th moment of the magnetisation, $m_k$,
the smallest volume which
can be reliably used grows quadratically with $k$.
At the transition point
the series of corrections to $m_k$ terminates at
the $1/V^{k-1}$ term. However, the coefficients in
this series grow extremely fast.
As a result, if the volumes are not very large,
the number of parameters required to describe the FSS
is equal to the number of moments studied.
This argues for using large volumes for FSS studies.

An alternative is to stay at intermediate volumes, and
accept the necessity of a large number of parameters,
but to perform many consistency checks. An example of
such a check is that the coefficients of the $1/V$
correction term for different moments are related in a
way independent of the Hamiltonian used. Good checks
necessarily require excellent statistics.

Finite-size effects in small volume systems
involve phenomena neglected in writing the expansion
in inverse powers of $V$. These are expected to
decrease exponentially with system size.
In order to control such errors, we propose
the study of cumulants. Within the approximations
which yield the power series corrections to the
moments, there are no finite-size effects in
properly normalised cumulants. Thus, observable
FSS effects in these quantities mark
the limits of the theory. Furthermore, these
cumulants are precisely the quantities whose
extraction motivates the study of the FSS theory.

This theory of finite-size scaling is developed
in section 2. We concentrate on the FSS theory for
the scaling of cumulants, and show how the theory
for moments follows from this. This extends the
results obtained in refs.~\cite{bk,bkm}. The nature of
the neglected corrections is discussed and a procedure
is developed to use simulation data to check whether
the theory is applicable. We comment on the relation
with FSS at a critical point.

In section 3 we present a test case where this theory
is numerically checked. This is done through
detailed numerical work
on the two-dimensional Ising model in its
low temperature ($1/\beta$)
phase along the line of zero external magnetic field.
This is a line of phase coexistence,
and is an obvious test-bed for
high-statistics numerical work on this problem.
The correlation length, $\xi$, can be tuned easily
by changing $\beta$. The statistical accuracy that can
be reached allows for a detailed study of the
finite-size effects. Valuable information about
the precise form of the FSS predictions is, moreover,
provided by existing analytical results. We study
lattice sizes $3<L/\xi<60$
in order to verify the expected
asymptotic size dependence and to find out how
large the system has to be before it sets in.
We end with a summary of our results in section 4.

\section{THEORY}

In this section we discuss the FSS equations for
moments of the order parameter.
For definiteness, we use the Ising notation, and
restrict the discussion to the case of
two symmetric coexisting phases. The generalisation
to arbitrary number of phases, not necessarily symmetric,
is straightforward, and shall be touched upon briefly.
{}From an asymptotic expansion of the free-energy
density in powers of $h$, for a $d$-dimensional system,
we obtain FSS expressions for moments of the magnetisation,
in powers of $L^{-d}=1/V$, at fixed
\beq
x\;=\;h V .
\label{expar}\eeq
The FSS theory for cumulants and moments are summarised
in eqs.~\ref{cumu} and \ref{mom} below.
We discuss the limits of validity of FSS models based
on single phase expansions
and write down expressions which we use in later sections.

\subsection{Formal results.}

At large $\beta$, the partition function in a periodic box
of size $V=L^d$, is approximated by \cite{bk}
\beq
Z(h,V)\,\approx\,Z_+(h,V)+Z_-(h,V)\ ,
\qquad Z_\pm(h,V)=e^{Vf_\pm(h)}\,.
\label{Zper}\eeq
Here $h=\beta H$ and $H$ is the external magnetic field.
Each term represents fluctuations about
one of the two coexisting phases.
The functions $f_{\pm}(h)$ are related
to the free-energy density
$${\bf f}(h)=\lim_{V\to\infty}{1\over V}\ln Z(h,V)$$
by ${\bf f}(h)=f_+(h)>f_-(h)$ if $h>0$, and a similiar equation with
an interchange of subscripts for $h<0$. This is supplemented by the
relation $f_+(0)=f_-(0)$.
Eq.~(\ref{Zper}) forms a good approximation for large $V$
because the omitted remainder is exponentially suppressed
in $L^{d-1}$.

The power corrections can be found by a formal
expansion of the functions $f_\pm(h)$ about $h=0$.
We write,
\beq
f_+(h)=\sum_{k=0}^\infty {a_k\over k!}h^k\,.
\label{as}\eeq
By symmetry, the corresponding coefficients for $f_-$ are
$(-1)^ka_k$. These series are asymptotic \cite{as}.
The first two coefficients have special names.
The coefficient $a_1=m_0$ is the spontaneous
magnetization and $\beta a_2=\chi$ the pure-phase
susceptibility.
The coefficients $a_k$ are related to cumulants
of the magnetisation, $\kappa_k^\pm$, in the phase
represented by the superscript, through the relation
\beq
\kappa_k^\pm(h=0,V)\;=\;(\pm)^ka_kV^{-(k-1)}\,.
\label{cumu}\eeq
Note that, apart from the usual neglected
exponential corrections, finite-size effects in
$\kappa_k^\pm$ are explicitly shown.
The normalised cumulants $\kappa_k V^{(k-1)}$
furnish estimates of $a_k$ on finite lattices.
The FSS theory for cumulants is now complete.

Formal manipulation of the expression in eq.~\ref{as}
yields FSS expressions
for $k$-th moment of the magnetisation, $m_k$, in the form
\beq
m_k(h={x\over V},V)=\sum_{j=0}^\infty {A_{k,j}(x)\over V^j}\,.
\label{gen}\eeq
Some care is required in the interpretation of these expressions.
In what follows, we shall focus on the behaviour
at the phase transition, $h=0$.
It is easy to see that $A_{k,j}(0)=0$ whenever $j\ge k$,
and the sum on the RHS of eq.~\ref{gen}
collapses to a finite number, $k$, of terms.
Unlike ref.~\cite{bk},
we shall include all these terms in our analysis.
In fact, terms beyond the second turn out to be
fairly crucial in understanding the structure of the
approximations.

Now we compute the precise form of the non-zero $A_{k,j}$'s.
It is convenient to start from the decomposition
\beq
m_k(h,V)=P_+(h,V)m_k^+(h,V) + P_-(h,V)m_k^-(h,V)\ ,
\label{decmom}\eeq
where $m_k^\pm(h,V)$ are the $k$th moment
with respect to the partial (single-phase)
distribution $Z_\pm(h,V)$, and are averaged with
weights $P_\pm(h,V)=Z_\pm(h,V)/Z(h,V)$.
The two weights are equal at $h=0$,
$$P_+(0,V)\;=\;P_-(0,V)\;=\;1/2.$$
The moments $m_k^\pm(h,V)$ are linear
combinations of the cumulants
$\kappa_j^\pm(h,V)$, $j=1,\ldots,k$, whose
size dependence have been shown in eq.~\ref{cumu}.
The size dependence of the moments, expressed
in terms of the constants $a_k$, can therefore be obtained
directly from the relations between moments and cumulants.
Odd moments vanish through symmetry. For even $k$, we have
\begin{eqnarray}
    m_k(0,V)\;&=&\;a_1^k
      + \Biggl\{\st_{k2}{k\choose 2}a_2 a_1^{k-2}\Biggr\}
                     {1\over V}\nonumber\\
 & &\qquad+ \Biggl\{\st_{k3}{k\choose 3}a_3a_1^{k-3}
      + \st_{k4}3{k\choose 4}a_2^2 a_1^{k-4}\Biggr\}
                     {1\over V^2}\nonumber\\
 & &\quad\qquad+ \cdots + a_k{1\over V^{k-1}}\,,
\label{mom}
\end{eqnarray}
where $\st_{ij}=1$ if $i\ge j$ and 0 otherwise.

This FSS formula for moments is for two symmetry-related phases.
It is easy to generalize this expression to the case of $N$
coexisting phases, not necessarily symmetric.
One starts from the scaling theory of single-phase cumulants.
This is always given by eq.~\ref{cumu}.
Then one uses well-known relations between cumulants and moments
to construct the latter in a single phase. Then the moments
actually needed are constructed by adding these with the proper
weights, in an appropriate generalisation of eq.~\ref{decmom}.
At the infinite-volume coexistence point, each phase occurs
with equal weight \cite{bk}. The $k$-th moment is thus the
arithmetic mean of the pure-phase moments. For superpositions
of asymmetric phases, all moments are in general non-zero. It is
also clear that at the infinite-volume coexistence point, in a
general model, for the $k$-th moment, the number of correction
terms in powers of $1/V$ cannot be larger than $k-1$.

We have presented explicit FSS expressions only for
the case $h=0$. It is also interesting to ask about
the behaviour away from the phase coexistence point
$h=0$. This can be studied by considering non-zero
values of $x$ in eq.~\ref{mom}.
The coefficients $A_{k,j}(x)$ can still be
expressed in terms of the parameters of the
free-energy expansion eq.~\ref{as}.
It is important to note that the FSS expansion is
given at fixed $x$ and not at fixed $h$. These two
variables are equal only for $x=h=0$.

\subsection{Nature of corrections.}

Corrections to the decomposition
of the partition function, eq.~\ref{Zper}, come from
exponentially suppressed contributions due to
phase mixtures and consequent interfaces.
Thus the results obtained should be valid only when these
contributions are negligible.
For this to be the case,
the barrier separating the minima of the free-energy
must be sufficiently high.
The bubble picture now yields a condition expressed
in terms of the surface tension $\sigma$ as
\beq
\sigma L^{d-1}\gg 1.\label{ineq1}\eeq
If Widom's relation, $\sigma\xi^{d-1}={\rm constant}$,
holds, then this is equivalent to the more
intuitively obvious inequality
\beq L\gg\xi.\label{ineq2}\eeq
For the two-dimensional Ising model, of course,
it is known from the
exact solution that $\sigma\xi=1/2$. However, due
to the fact that we are dealing with an asymptotic
expansion of the free energy,
these inequalities must be interpreted with care.

The FSS theory is based on the
approximation of eq.~\ref{Zper}.
As mentioned before, the FSS theory for the cumulants
(eq.~\ref{cumu}) thus contains corrections which decrease
exponentially with a power of $L$ at large $L$. However,
at fixed $L$, cumulants of higher order are more sensitive
to the neglected portions of the partition function.
Numerical work indicates that the minimum volume
$V_k^{(0)}$ at which eq.~\ref{cumu} is true for the
$k$-th cumulant increases with $k$. Thus one requires
larger system sizes $L/\xi$ for larger $k$.

Corrections to the FSS theory of the moments (eq.~\ref{mom})
are of two types. The finite-volume expansion for moments is
written in terms of the cumulants. At the volume where the
FSS theory of a cumulant breaks down, the FSS expansion of
all moments which involve this cumulant must also break down.
Thus the expansion of eq.~\ref{mom} for $m_k$ cannot be
expected to be correct for volumes smaller than $V_k^{(0)}$.

A second effect, which may not be totally unrelated,
concerns the expansion parameter in the series in
eq.~\ref{mom}. When the system volume is very small,
an expansion in $1/V$ cannot be meaningful. An estimate
of the volume below which one should not use this
expansion is given by the size, $V_k$, at which the
highest term is dominant in the FSS series.
The dominance of the highest
cumulant further indicates that the neglected portion
of the partition function may start becoming important.
We cannot say, a priori, whether $V_k$ or $V_k^{(0)}$
provides a more stringent bound on the applicability of
the FSS theory given in the previous subsection.

\begin{figure}[htbp]
\vskip 6truecm
\includegraphics{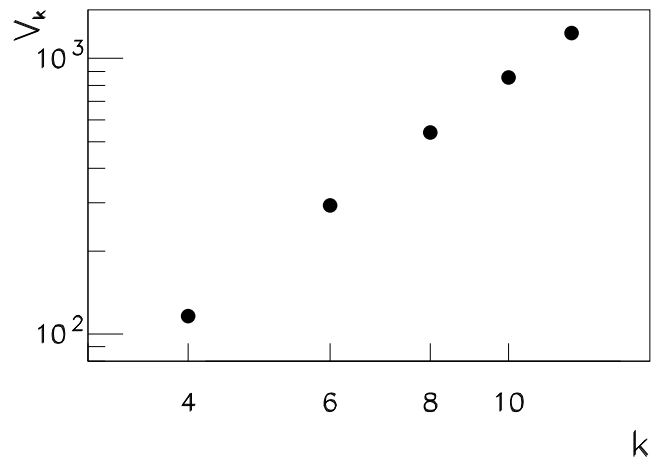}
\caption{The dependence of $V_k$ on $k$.}
\label{fig:vk}
\end{figure}

Unlike $V_k^{(0)}$, it is possible to analyse $V_k$ in
further detail. We define this as the volume at which
the absolute value of the term in $1/V^{k-1}$ becomes
equal to the term in $1/V$ in the expansion given in
eq.~\ref{mom}. Thus,
\beq
V_k= {1\over a_1}
\Biggl({2|a_k|\over k(k-1)a_2}\Biggr)^{1\over k-2}\ .
\label{vk}\eeq
Using the results of ref.~\cite{bakerkim},
we find that $V_k$ grows approximately quadratically with $k$.
This is shown in figure~\ref{fig:vk}. For volumes of the order
of $V_k$, the highest correction dominates in eq.~\ref{mom}.
This indicates that the expansion parameter $1/V$ is
not small. Hence, we cannot then expect this expression
to provide a good description of the data.
Moreover, for large $k$ at fixed volume,
the second-order correction is larger than
the leading term, indicating that even a truncated
sum does not make sense in the limit of large $k$ unless
$V$ is made larger. Eq.~\ref{mom} should therefore be a
good approximation on a given volume only if
the order of the moment is not too high.

These arguments can be restated in the following way.
Since the expansion of the free energy is asymptotic,
for fixed $h$ there is a value of $n$, say $N_o$,
which minimises the truncation error
\beq
\left| f(h)-\sum_{k=0}^n {a_k\over k!} h^k\right| <
C_{n+1} |h|^{n+1} \ .
\label{tr}\eeq
The optimal choice, $N_o(h)$,
depends on $h$ and becomes large as $h$ gets small.
Adding further terms to the sum makes the approximation
worse. This has implications for the study of cumulants
and moments of varying order. To study the $k$th
cumulant or moment, we clearly need to consider
a partial sum of at least $k$ terms. For this
partial sum to make sense,
$h$ must be small enough that
$N_o(h)>k$. For the finite-size expansion
at fixed $x\ne 0$, this indeed implies that $L$ must
be taken larger as the order $k$ increases. We
shall see below that the same behaviour holds
at $x=0$.

\subsection{Near criticality.}

The normalised cumulants $a_k$ are implicitly dependent on
the temperature. So far we have assumed this to be fixed.
We can vary the strength of the phase transition
by changing the temperature. As $\beta\to\beta_c^+$,
the transition weakens, and approaches criticality.
It is interesting to see how the FSS theory then approaches
the standard theory at a second-order transition.

The temperature dependence of the normalised cumulants near
$\beta_c$ define the critical exponents. Thus, in terms of
the reduced temperature $t=(T_c-T)/T_c$, we obtain
\beq
a_k\sim t^{\nu( d-k y_H)}.
\label{scum}
\eeq
Here, $\nu$ is the correlation-length exponent and
$y_H$ is the field exponent. The scaling of moments is easily
derived. In eq.~\ref{mom}, the $j$th correction to the $k$th
moment scales as
\beq
{A_{k,j}(0)\over V^j}
   \sim t^{k\beta}\Biggl({L\over\xi}\Biggr)^{-jd}
\label{scal}\eeq
where $\beta=\nu(d-y_H)$ is the magnetization exponent.
For a given $k$, the first factor is common to all the terms
in the expansion. This yields the infinite volume behaviour.
The $L$ dependence comes from the second factor and it enters
explicitly only through the ratio $L/\xi$. In applications to
the analysis of Monte Carlo data, at fixed $L$, higher
corrections are more important the weaker the transition.
To the extent that the temperature dependence of $a_k$ is
given by the critical point formula in eq.~\ref{scum},
the functional dependence of the moments on $L/\xi$
is independent of the coupling, and hence,
of the strength of the transition.

In a similiar way one can investigate the dependence of $V_k$
on the temperature, as one approaches criticality. Using the
scaling laws of eq.~\ref{scum} along with the definition of
$V_k$ given in eq.~\ref{vk}, we find
\beq
V_k(t)\;=\;{\cal V}_k \xi^d(t).
\label{scvk}\eeq
Here ${\cal V}_k$ is independent of $t$ (to the extent that
eq.~\ref{scum} contains the $t$-dependence of $a_k$) and we
have taken $\xi\sim t^{-\nu}$. For each $k$, as one approaches
the critical point, the volumes over which the FSS theory is
valid must grow as $\xi^d$. However, the ratio of $V_k$ for
different $k$ is independent of the temperature. Hence,
arbitrarily close to the critical point, the correct scaling
behaviour of the higher moments can be seen only on larger
volumes.

\subsection{Numerical implications.}

The main consequence of the FSS theory is that it
allows the extraction of bulk thermodynamic
quantities from experiments with small systems.
The physical quantities required from such experiments
are the single-phase cumulants. The primary consistency
test of the FSS theory has been in a good description
of finite-size effects for moments with an unique set
of cumulants. Usually only the first correction term
in eq.~\ref{mom} has been retained for such a check.
It is easy to extend this procedure, and keep as many
terms as necessary. The restrictions on this procedure
have been discussed in detail already.

We also propose a second procedure. The data can be used
to extract normalised single-phase cumulants, $a_k(L)$.
Their constancy with changing $L$ signals the
applicability of the theory; size-dependence outside errors
indicates that the lattice sizes are too small for the
FSS theory to be applicable. The moments can be used as
subsidiary checks of the FSS theory. This procedure depends
on our ability to extract $a_k(L)$ directly from the data.

We suggest a definition of these cumulants by introducing
a cut, $m^*$, between the two peaks in the probability
distribution. In the part retained, say, $m>m^*$,
we expect contributions from the disfavoured
phase (as defined by the decomposition of eq.~\ref{Zper})
to be exponentially suppressed in $V$. Phase mixtures
must also be suppressed in a similiar fashion.
Up to exponentially small errors, we therefore expect
\beq
a_k\,=\,V^{k-1}\langle m^k\rangle^*_c\,,
\label{cum}\eeq
where the $\langle\cdots\rangle^*$
indicates expectation values over $m>m^*$.

The choice of $m^*$ is not very complicated. It should
clearly satisfy $|m^*|<<m_0$, and, provided the
free-energy density near the minimum is flat, can be
chosen anywhere in this interval. The origin of the
exponential bound on the error is, of course,
intimately related to the existence of this flat region.
Choosing $m^*$ outside this interval
would increase the constant in front of the
exponential that bounds the errors.
Note also that the measurement of
$\langle m^k\rangle^*_c$ rapidly
becomes more difficult with increasing order $k$,
since it is $O(L^{-(k-1)d})$ and is the result of
cancellations between $O(1)$ numbers.
This is not a problem, since the cumulants which
are difficult to measure on a given volume should
certainly to be ignored in the FSS.
We demonstrate the feasibility of such measurements in
the next section.


\subsection{Other observables.}

In the FSS theory of cumulants and moments, constructed
here, the single-phase
contribution to the partition function is of the form
$Z_i(h,V)=e^{Vf_i(h)}$. It is of interest to construct the
corresponding probability distribution for
the order parameter, $P_i(m,V)$, in the phase labelled
by $i$. In order to obtain
the first two terms in the FSS expansion of moments
we have seen that it is
sufficient to consider a quadratic
approximation of $f(h)$, $\hat f_i(h)=a_0+a_1h+a_2h^2/2$.
By inverse Laplace transform, the corresponding
probability distribution is a Gaussian,
\beq
\hat P_i(m,V)\,=\,{\rm cst}\times
\exp\Biggl(-V{(m-a_1)^2\over 2a_2}\Biggr)\ .
\label{gauss}\eeq
This distribution, used in refs.~\cite{bl,clb},
therefore produces the same first two terms
in the FSS expansion of moments.

The situation can be quite different for other quantities.
As an example, consider the position of the maximum in the
probability distribution $P_i(m,V)$. This is equivalent to
finding the minimum of the effective action
\beq
s_i(m,V)=-{1\over V}\ln P_i(m,V).
\label{effpot}\eeq
The Gaussian approximation for $P_i(m,V)$
(eq.~\ref{gauss}) would predict no finite-size shift in
this quantity. On the contrary, such a shift has
been seen \cite{lk}. A saddle point expansion shows
that the minimum of $s_i$ is at
\beq
m\;=\;a_1-{a_3\over 2a_2}{1\over V}+\ldots\ .
\label{shift}\eeq
Since $a_3\ne0$ in general, we expect a finite-size shift of
order $1/V$.

In a study of two-dimensional Potts models \cite{lk}, it has
been claimed that the shift of the minimum of the effective
potential scales as $1/L$. It would be interesting to extend
these computations to volumes where the approximation in
eq.~\ref{Zper} is valid to good accuracy in order to check
whether there is a crossover to a $1/V$ behaviour, as
predicted here.

\section{TESTS}

{\begin{table*}[t]
\setlength{\tabcolsep}{1.5pc}
\caption{The couplings, range of lattice sizes and statistics
for all the runs. Also indicated are the exact values of the
spontaneous magentisation and correlation lengths at these
couplings.}
\label{tab:runs}
\begin{tabular}{rrrrr}
\hline
 $\beta$ & $m_0$ & $\xi$ & $L$ & $N_{sw}$ \\
\hline
  $0.4670$ & $0.84247$ & $4.83$ & 16--60 & 8 $\times10^6$ \\
  $0.4750$ & $0.86578$ & $3.73$ & 16--128 & 3.6--10 $\times10^6$ \\
  $0.5000$ & $0.91132$ & $2.19$ & 16--128 & 2--10 $\times10^6$ \\
  $0.5500$ & $0.95394$ & $1.22$ & 16--64 & 2--4 $\times10^6$ \\
\hline
\end{tabular}
\end{table*}}

In this section we present some details of a numerical test
of the FSS theory developed in the previous section.
These involve data collected in simulations of the
two-dimensional Ising model in its ordered phase.
We concentrate on the scaling of moments
and extraction of pure-phase cumulants of the magnetisation.
We study the efficacy of different methods for the
extraction of the second cumulant, {\it i.e.\/}, the susceptibility
in a pure phase. Extensive simulations were performed at four
couplings. One was chosen such that $u/u_c=0.9$,
where $u=\exp(-4\beta)$. Since
$u_c=3-2\sqrt{2}$, this implies $\beta\approx0.467$.
This choice was motivated by the fact that several
of the cumulants are known to high precision from power-series
expansions \cite{bakerkim} at this coupling.

We used the Swendson-Wang cluster algorithm to
simulate $L^2$ lattices with periodic boundary conditions.
Measurements were separated
by 10 iterations. Expectation values and errors
were estimated through a jack-knife procedure.
The couplings at which simulations were performed
are listed in Table~\ref{tab:runs}, along with some of the
relevant details of the runs. When a range of statistics is
metioned, then the more extensive statistics were
taken on the larger lattices. At all couplings and
volumes used in our study, the relative errors on
the second moment of the magnetisation was kept below
$2\times10^{-4}$, and that for the eighth moment
below $10^{-3}$.

The data at the smallest $\beta$ have been used for
a straightforward extraction of cumulants using the method
of cutting the data discussed at the end of the previous
section. We present comparisons of these measurements with the
results of series expansions. The values of the cumulants
were used to check the FSS theory by comparing data on the moments
to the complete FSS expansion in powers of $1/V$.

\begin{figure}[bthp]
\vskip 6.5truecm
\includegraphics{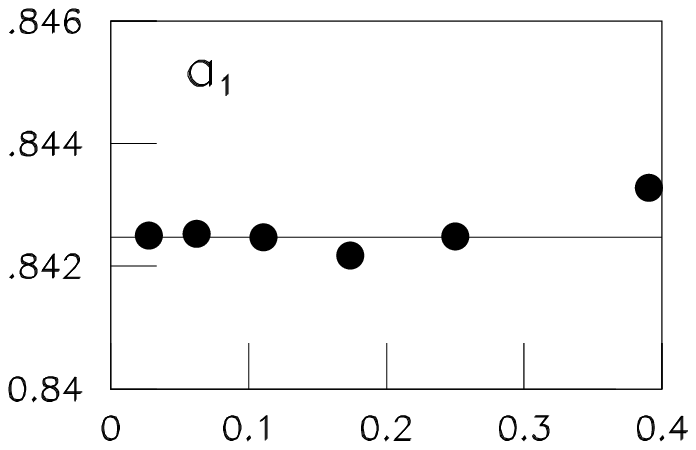}
\includegraphics{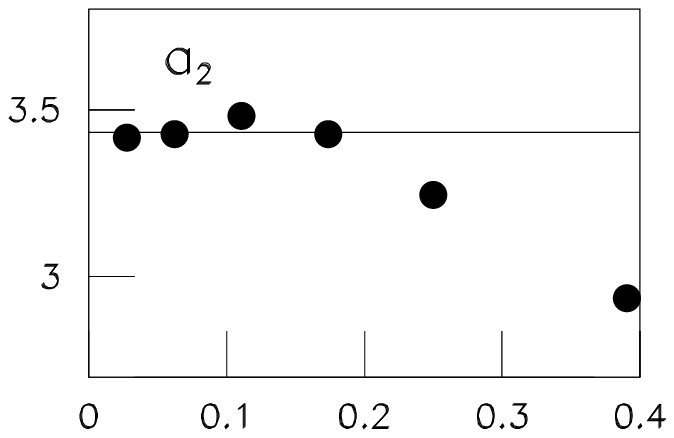}
\includegraphics{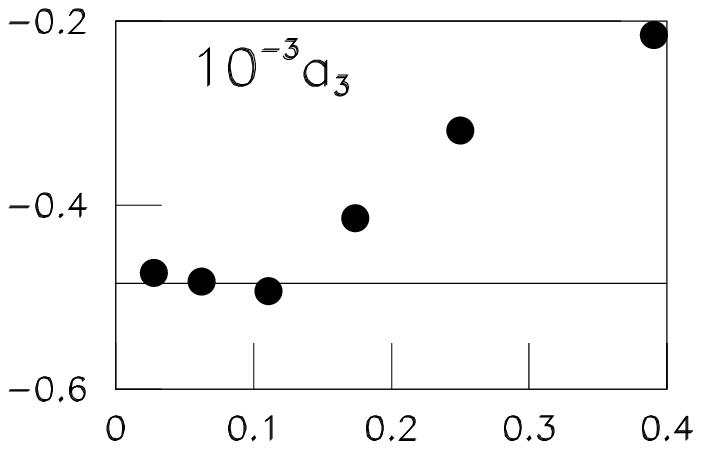}
\includegraphics{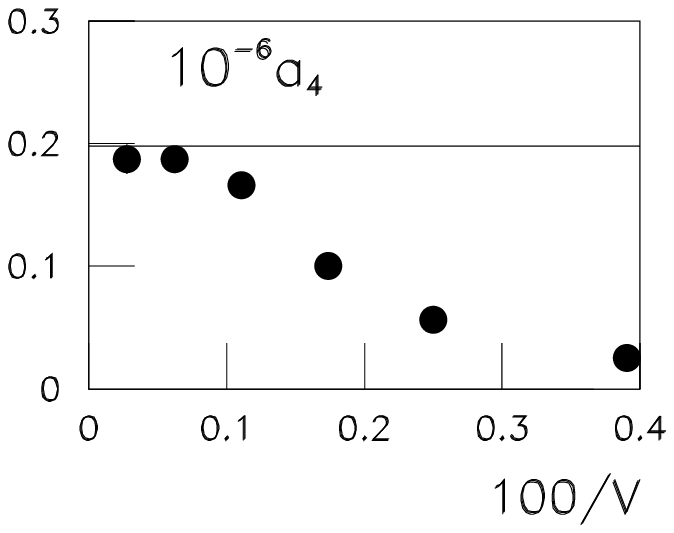}
\caption{Normalised single-phase cumulants $a_k$ against $1/V$.
  The error bars are smaller than the size of the symbols. The
  horizontal lines show the values from series expansions.}
\label{fig:cumk}
\end{figure}

The data at the larger couplings were taken
on a wider range of lattice sizes.
This data has been used for a direct analysis of the scaling
of moments. The aim is to find the range of
lattice sizes which is sufficiently large that the
leading correction terms in the FSS
expansion are sufficient to describe the
data. The scaling of data on these lattices then allow us
to extract the pure-phase susceptibility. Thus, we illustrate
two different methods to extract physics from finite-size
scaling at phase coexistence.

\subsection{The coupling $u=0.9u_c$.}

\begin{figure}[thbp]
\vskip 6.2truecm
\includegraphics{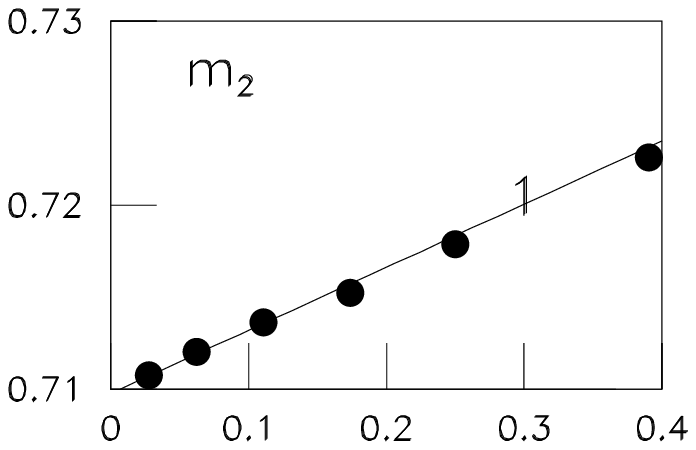}
\includegraphics{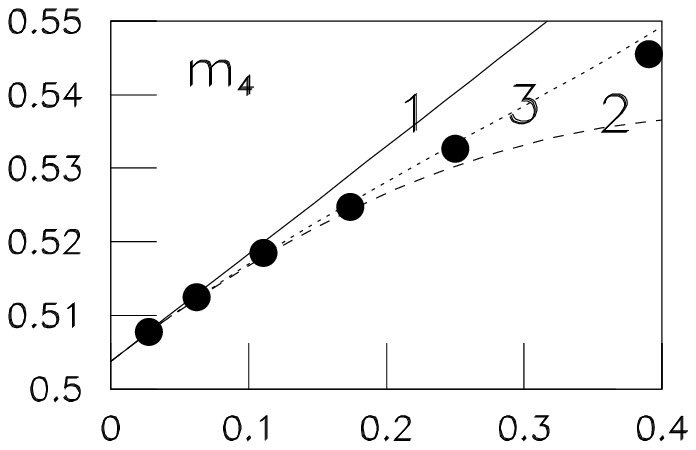}
\includegraphics{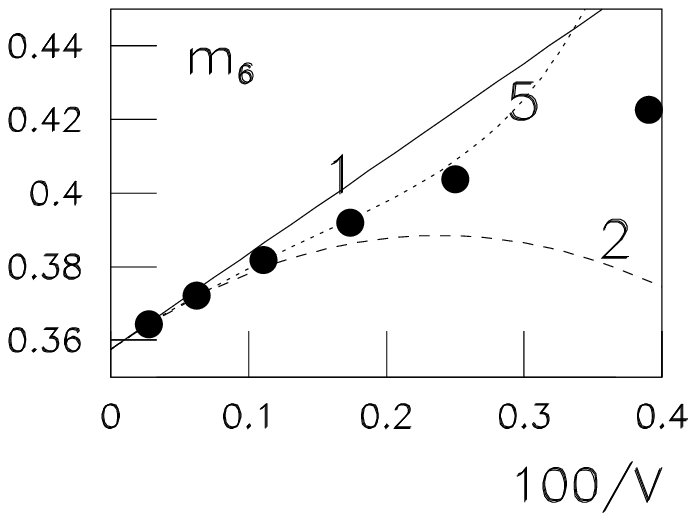}
\includegraphics{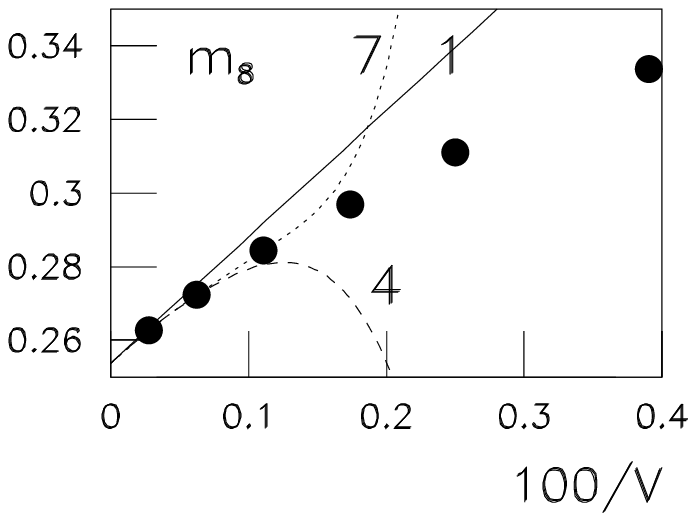}
\caption{Moments $\langle m^k\rangle$ against $1/V$. The error
  bars are smaller than the size of the symbols. The curve
  labelled $j$ is the prediction obtained by including terms
  of order up to $1/V^j$. To avoid clutter, no more than three
  curves have been shown for each moment.}
\label{fig:momk}
\end{figure}

The first four normalized single-phase cumulants $a_k$
(see eq.~\ref{cum}) are shown in figure~\ref{fig:cumk}.
They were obtained using the cut $m^*=0$, and by folding
data for negative magnetisations into the interval $m\ge0$.
In the FSS theory we expect these cumulants to approach
a constant. Clearly, the smallest lattices lie outside the
range of validity of the theory. It is clear from the
figure that $V_k^{(0)}$ increases with $k$. For larger
$V$ the cumulants are consistent with their known asymptotic
values \cite{bakerkim} within errors. The
first two cumulants can be determined with an error of
less than 3\% on lattices with $L/\xi>5$. This accuracy
is sufficient for this measurement to be useful in the
FSS analysis of moments. The rapid
approach to the large volume limit is extremely useful
for practical determination of the cumulants.
However this also means that
a numerical investigation of the neglected
terms in the partition function is not feasible
without much higher statistics.

The known values of $a_k$ \cite{bakerkim} can be used for
the scaling of the moments $m_k$ using the formula
in eq.~\ref{mom}. These predictions are shown along with
the data in figure~\ref{fig:momk}. The previously studied
$1/V$ correction corresponds to the straight line and
seems to give a good approximation at large volumes.
Note that the FSS theory predicts no higher term for
the scaling of the second moment. The data is completely
consistent with this expectation for $L/\xi$ as low as 4.
Consistent with this, we found that a fit to the data on
$m_2$ for $L/\xi\ge4$ correctly yields the values of the
first two cumulants.

The exact curves show that the fourth and higher moments
definitely require terms beyond the leading $1/V$
correction for a description of the data for
$L/\xi\lsim8$. It is interesting to note that an
attempt to fit a $1/V$ correction term to the data on
$m_4$ for $L/\xi>6$ produces a good fit by a $\chi^2$
measure. However, the values of the first two cumulants
so obtained are wrong by 7--10$\sigma$. This state of
affairs is similiar to the situation reported for
the two-dimensional 10-state Potts model \cite{blm}.

The volume dependence of the sixth and eighth moments
cannot be fitted purely by the $1/V$ term for
$L/\xi\ge6$. Note also that the exact prediction
using the FSS model of eq.~\ref{mom} deviates strongly
from the data at increasingly larger volumes as
$k$ grows.

\begin{figure}[bhtp]
\vskip 6.5truecm
\includegraphics{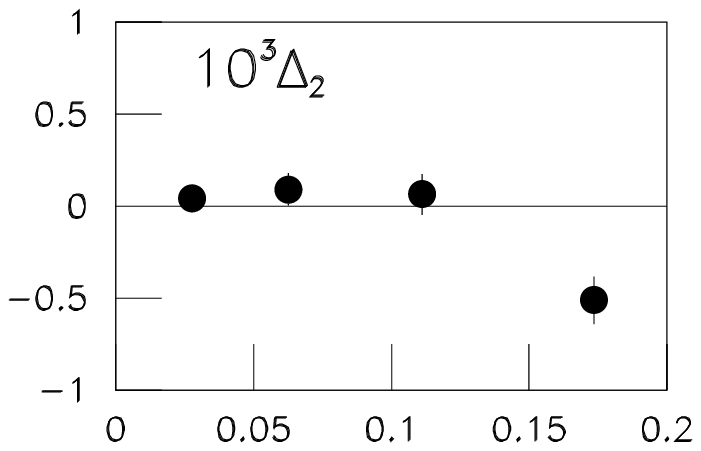}
\includegraphics{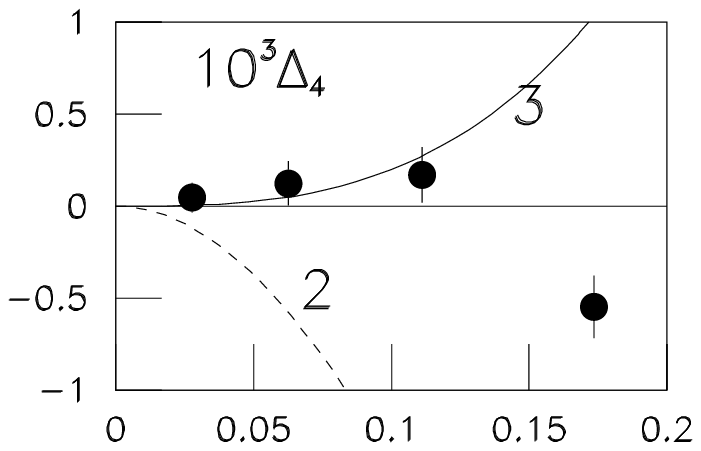}
\includegraphics{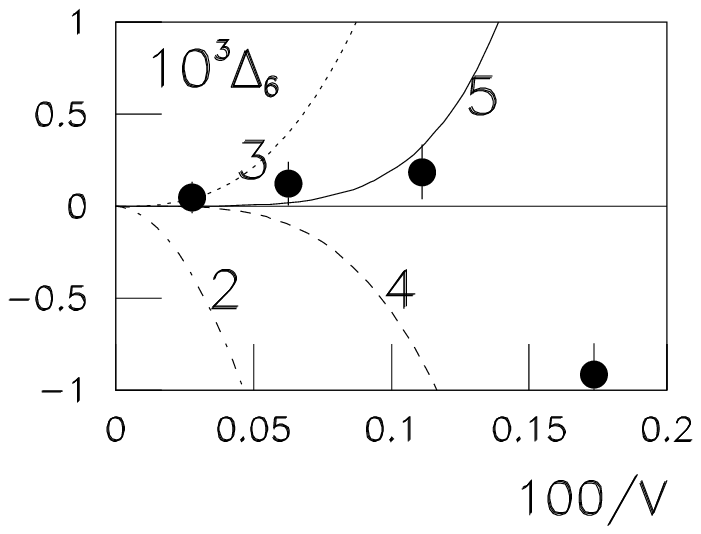}
\includegraphics{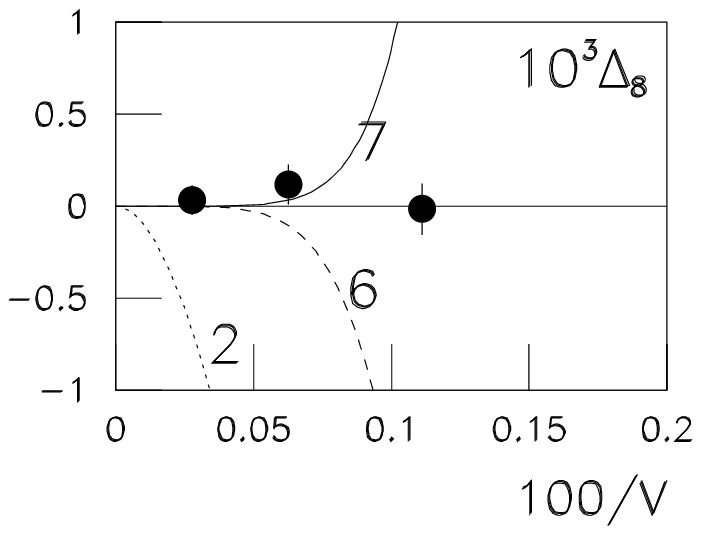}
\caption{The deviations $\Delta_k$ against $1/V$. For each
moment $k$, the residue $R_j$ is labelled by the number $j$.
To avoid clutter, only 3 of the curves are shown for $k=8$.}
\label{fig:delk}
\end{figure}

For a more detailed assessment of this volume dependence,
we studied the difference between the predictions of
eq.~\ref{mom} and the measured values of the moments
\beq
\Delta_k\,=\,m_k({\rm meas})-m_k({\rm pred})\,.
\label{delta}\eeq
These values are plotted in figure~\ref{fig:delk} as a
function of $1/V$. Also plotted are the values of the
residues
\beq
R_j^{(k)}\;=\;\sum_{i=j}^{k-1}{A_{k,i}(0)\over V^i}\,.
\label{curves}\eeq
When analysing FSS for $m_k$ in terms of a series
truncated at the $1/V^j$ term, these residues are
the sum of the neglected terms. It is clear that
at small volumes correction terms beyond $1/V$
are quite important.

Another striking feature that emerges is that,
as $V$ decreases, there seems to be a fairly abrupt threshold
at which the agreement between the $\Delta_k$ and the residues
disappears. This threshold volume increases with $k$, and is
a little larger than $V_k$. The data seems to indicate that
the FSS theory of moments is valid for volumes $V>V_k^{(0)}$.
The limit $V_k$ is less stringent for this case.

That there is a lower limit to the volumes for which
the FSS theory is applicable is also shown in
figure~\ref{fig:ldelk}. Here we plot $\ln|\Delta_k|$
against $L$. For $L/\xi\sim3$,
it turns out that the measured values of the higher
moments are much smaller than the value predicted by
the series in eq.~\ref{mom}. As a result $\ln|\Delta_k|$
is large in this region. Since the highest term in the
FSS series tends to dominate at small volumes, for very
small $L$ we should see the trivial behaviour
$\ln|\Delta_k|\sim(k-1)d\ln L$. At much larger values of
$L/\xi$, $m_k$ should have a good description in terms
of the FSS series. In this region we expect small values
of $\Delta_k$, going to zero roughly as
$\ln|\Delta_k|\sim(d-1)L$. The data for $k=10$ clearly
exhibits a cross-over from a regime of large $\Delta_k$
to small values; the latter being reached only on the
two largest lattices. In this region, the data is too
noisy to characterise the approach to zero, and hence
the nature of the neglected error terms.


\begin{figure}[htb]
\vskip 6truecm
\includegraphics{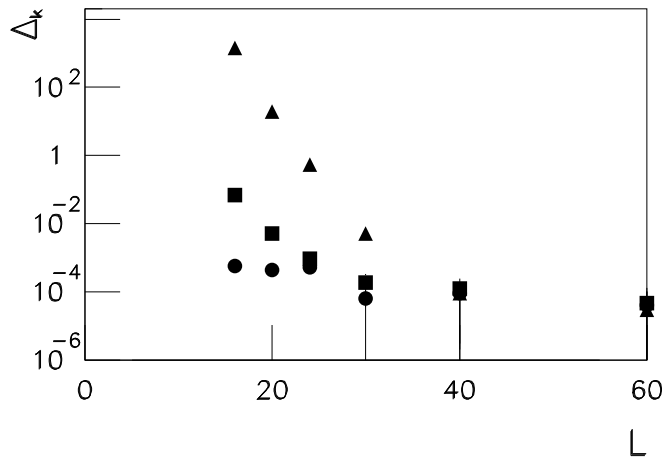}
\caption{The deviations $\Delta_k$ for $k=2$ (circles), 6
(squares) and 10 (triangles), plotted against $L$.}
\label{fig:ldelk}
\end{figure}

\subsection{Other couplings.}

The leading terms of the FSS theory at $h=0$ for the
scaling of moments are
\beq
m_{2k}\;=\; m_0^{2k} + {\alpha_k\over V}
   + {\beta_k\over V^2} + {\gamma_k\over V^3},
\label{momex}
\eeq
where eq.~\ref{mom} contains expressions for
$\alpha_k$ and $\beta_k$ in terms of the cumulants.
The coefficient $\alpha_k$ is related to the pure
phase susceptibility, $\chi$, and the spontaneous
magnetisation, $m_0$, by the relation
\beq
\alpha_k\;=\;(2k-1)k m_0^{2k-2}{\chi\over\beta}.
\label{momexp}
\eeq
Thus, for sufficiently large volumes, analysis of
different moments should yield consistent values of
$m_0$ and $\chi$. At the coupling $\beta=0.475$,
we have data on lattices with $4.3\le L/\xi\le34.3$.
For lattice sizes which are not too large, we find
that it is possible to describe the data fairly well
using only the four terms shown in eq.~\ref{momex}.
In figure~\ref{fig:momf} we show fits to the first four
even moments of the magnetisation for $L/\xi\ge10$.
At the two remaining couplings, data for lattices of size
$L/\xi\ge10$ are also very well described by the
same number of terms.

\begin{figure}[hbt]
\vskip 6.2truecm
\includegraphics{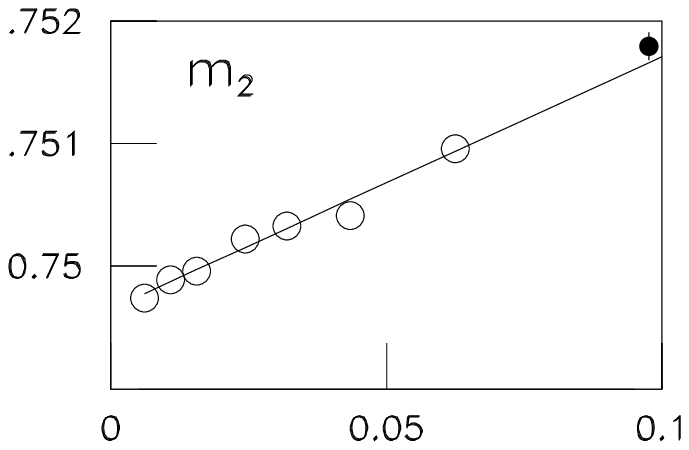}
\includegraphics{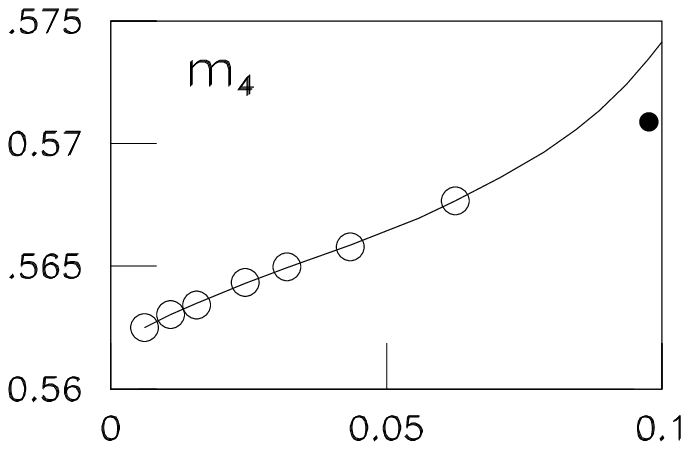}
\includegraphics{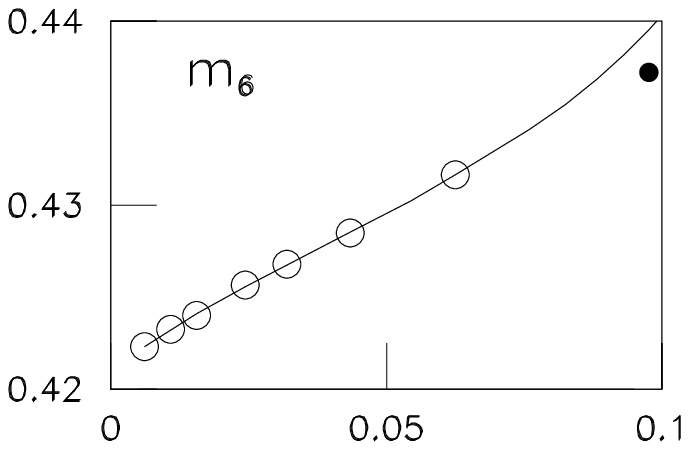}
\includegraphics{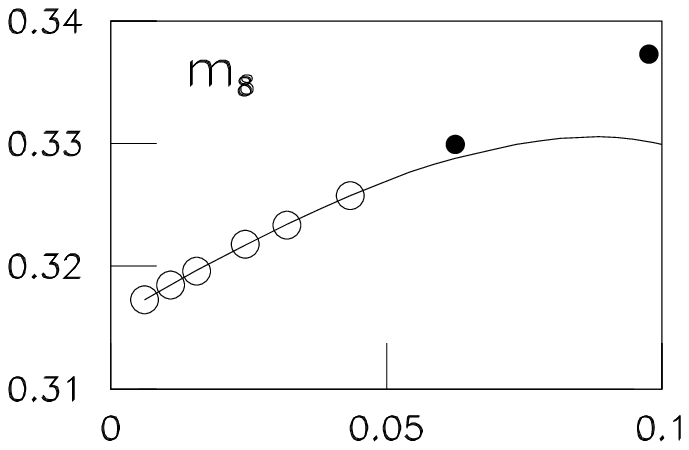}
\caption{The moments $m_{2k}$ at $\beta=0.475$, for $1\le k\le4$,
plotted against $1/L^2$.
 The curves show the best fits to the forms
described in the text. The open circles indicate the data to
which these fits have been made.}
\label{fig:momf}
\end{figure}

{\begin{table*}[t]
\setlength{\tabcolsep}{1.5pc}
\caption{The spontaneous magentisation extracted from fits
to $m_{2k}$ at the couplings shown. The errors on the final
digit are indicated by the numbers in brackets.}
\label{tab:spon}
\begin{tabular}{rrrrr}
\hline
 $\beta$ & $k=1$ & $k=2$ & $k=3$ & $k=4$ \\
\hline
  $0.4750$ & $0.86583(2)$ & $0.86581(3)$ & $0.86581(3)$ & $0.8657(3)$ \\
  $0.5000$ & $0.91135(2)$ & $0.91135(2)$ & $0.91135(2)$ & $0.91136(2)$ \\
  $0.5500$ & $0.95395(3)$ & $0.95400(4)$ & $0.95400(4)$ & $0.95400(4)$ \\
\hline
\end{tabular}
\end{table*}}

These fits yield independent estimates of $m_0$ from
the extrapolated infinite volume value of each moment.
These values, extracted from the first four even moments
at three different couplings, are given in Table
\ref{tab:spon}. They are completely consistent with the
known exact results
listed in Table \ref{tab:runs}, and with each other.
Note that equally good values for $m_0$ can be obtained
by using only the first two terms of eq.~\ref{momex} to
describe the data on lattices with $L/\xi\ge15$.

The same fits allow for a second check of consistency.
If we denote by $\chi^{(k)}$ the value of the pure-phase
susceptibility extracted from fits to the data on $m_{2k}$, and
by $m_0^{(k)}$ that of the spontaneous magnetisation, then
this gives a prediction for the coefficient
\beq
\alpha_l^{(k)}\;=\;(2k-1)k (m_0^{(k)})^{2k-2}
       {\chi^{(k)}\over\beta}.
\label{pred}
\eeq
Consistency demands that for any $l$, the predicted values of
$\alpha_l^{(k)}$ (all $k$) should be consistent within errors with
the value of $\alpha_l$ obtained by a direct fit to the data. That
this is true is shown by the values listed in Table \ref{tab:consi}.

\begin{figure}[bht]
\vskip 6truecm
\includegraphics{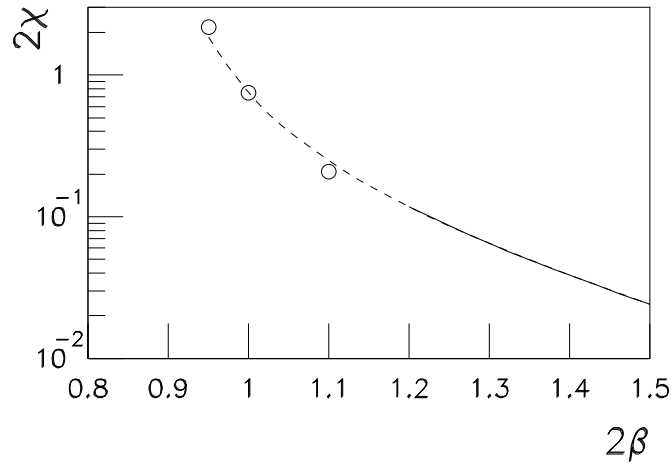}
\caption{The pure phase susceptibility $\overline\chi$ at
three different couplings, extracted from FSS of data on
the moments of the magnetisation. The lines are obtained
from series expansions as explained in the text.}
\label{fig:chi}
\end{figure}

This check indicates that the data requires the retention
of terms beyond $1/V$ even for lattice sizes as large as $L/\xi=15$
or more. In an attempt to analyse these large lattices with only
the first two terms in eq.~\ref{momex}, we found that the values
of $\alpha_l^{(k)}$ for each $l$ systematically decreased with
increasing $k$. This is consistent with the fact that the next
coefficient, $\beta_k$, is negative. This effect was not very
strong; at the $3\sigma$ level there was no systematic effect.
Clearly, with half as much data, we would have seen no effect.
Inclusion of the $1/V^2$ term in the analysis reversed the trend;
again the effect was weak, and could be neglected at
the $3\sigma$ level. The inclusion of the $1/V^4$ term removed
all trends within errors. This also increased the stability of
the value of $\chi$ extracted from $\alpha_k$.

{\begin{table*}[t]
\setlength{\tabcolsep}{1.5pc}
\caption{Tests of the consistency of the first order FSS fits
for $\beta=0.475$. The values of $\alpha_l^{(k)}$ are listed.
The errors are obtained using the full covariance matrix of
the fitted parameters. The diagonal entries, $\alpha_l$, are
from direct fits to $m_{2l}$.}
\label{tab:consi}
\begin{tabular}{rrrrr}
\hline
 $k$ & $l=1$ & $l=2$ & $l=3$ & $l=4$ \\
\hline
  $1$ & $2.1(1)$ & & & \\
  $2$ & $9.3(5)$ & $9.9(6)$ & & \\
  $3$ & $17(1)$ & $19(1)$ & $18.3(7)$ & \\
  $4$ & $24(1)$ & $26(2)$ & $26(1)$ & 26.9(9) \\
\hline
\end{tabular}
\end{table*}}

This consistency allows us to extract a stable and unique
value for $\chi$ from each set of data.
These are shown in figure~\ref{fig:chi}.
We have also estimated the pure
phase susceptibilities using a low-temperature
expansion to order 9 in $u$ \cite{domb}.
This is sufficient for couplings at which the correlation
length is less than 3.
For the smaller couplings this series is extrapolated using
standard techniques \cite{gauntgutman}. The results obtained
from the FSS measurements are consistent with the series
results.

\subsection{Approach to criticality.}

We have seen that the asymptotic size dependence
at a fixed temperature is well described by the
FSS theory for a first-order phase transition.
If this temperature is near the critical end point,
then we would at the same time expect FSS
theory for a second-order phase transition to apply.
In Section 2, we saw that these requirements are
consistent. The formula for the $k$th moment, as obtained
from FSS for a first-order phase transition, can, in
the vicinity of $\beta_c$, be written in the form
\beq
{m_k(\beta,h=0,L)\over m_o^k}\,=\,1+f(L/\xi) \ ,
\label{temp}\eeq
where the finite-size correction $f$ depends
only on the ratio $L/\xi$.
In figure~\ref{fig:crit} we have
plotted the fourth moment against
$(\xi/L)^2$ for two different temperatures. Eq.~\ref{temp}
implies that two data
sets should fall on a common curve.
The figure confirms that the temperature dependence
to a good approximation is described by eq.~\ref{temp}.

\begin{figure}[hbt]
\vskip 6truecm
\includegraphics{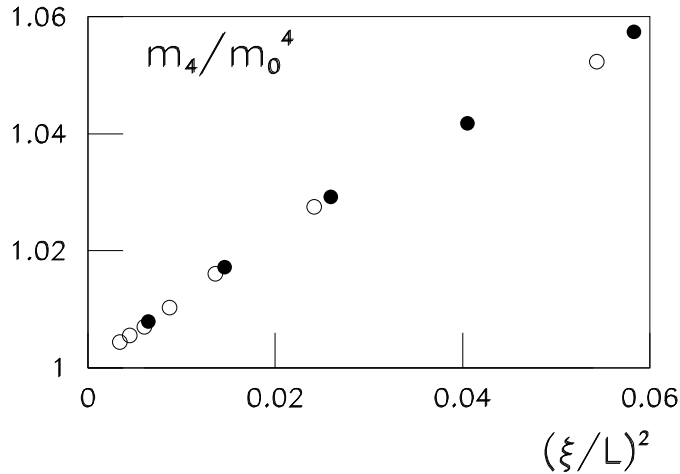}
\caption{The fourth moment of the magnetisation at two
different temperetaures, $u/u_c=0.9$ (filled circles) and
$2\beta=0.95$ (open circles).}
\label{fig:crit}
\end{figure}

\section{CONCLUSIONS}

A finite-size scaling theory at phase coexistence,
based on a formal expansion of the free energy density
in terms of an external field or coupling, gives a very
good description of observations when the system size
$L$ is large enough. Data taken in the low-temperature
phase of the Ising model indicates that the scaling of
lower moments of the magnetisation for $L/\xi\gsim6$
is completely consistent with the theory, once all terms
in the expansion have been taken into account.
The second moment seems to be completely
consistent with the theory for lattice sizes as small
as $3\xi$. Higher order moments generally require larger
volumes for the theory to be applicable. Several correction
terms seem to be necessary even for lattices as large as
$L/\xi\ge15$.

The relative importance of the higher correction
terms is, of course, dependent on the actual values of
$a_k$, and hence on the Hamiltonian studied. For the
Ising model, these cumulants decrease with increasing
$\beta$. Thus, when the temperature is small enough,
and $\xi$ is smaller than one lattice spacing,
the correction term of order $1/V$ suffices even for
rather small $V$. This is also true for two dimensional
Potts models with sufficiently large number of states.

In the numerical study of a phase transistion a first task
is to determine the order of the transition.
As is well known, this can be done, for example,
by examining the size dependence of the
susceptibility or specific heat. A linear divergence
of the response function with increasing volume
signals the first-order nature of the phase
transition. Clearly, this requires a
FSS study, but it should be noticed one is here
investigating a leading-order effect.
The actual usefulness of the full FSS theory in this context
is therefore not obvious \cite{bnb}.

Having established the first-order nature of a
phase transition, the primary physical
quantities are cumulants,
{\sl i.e.\/}, derivatives of the free-energy, for
the coexisting phases. These quantities represent
higher order effects in the system size. It is an
important feature of the FSS theory that the
finite-size corrections
to these cumulants are predicted to be exponentially
small. This suggests a direct measurement of
them. For this purpose
we have tested the method of pure-phase cuts.
We found that the (lower) cumulants can be obtained in this
way with reasonable accuracy
on small volume systems ($8\le L/\xi\le12$).
Indeed, we have seen a rapid approach to the
infinite volume results.

A further role of the FSS theory
is to provide the possibility to perform important
consistency checks.
Once the cumulants are known, the FSS
theory predicts the behaviour of the moments, which can be
easily tested. Alternatively,
when good data on intermediate volumes ($10\le L/\xi\le30$)
are available, then direct FSS fits can be used to extract
the cumulants and thus obtain independent estimates of these.
Our experience suggests that the direct extraction
through single-phase cuts is the easier of the two methods.

We have shown how the FSS theory for coexisting phases goes
over smoothly into the standard FSS theory for a critical
system. We note further that the expansion used in developing
the FSS relations is an asymptotic expansion. We have worked
at the infinite volume coexistence point, where the expansions
terminate at a finite number of terms, and the corrections to
this result are exponentially small.

Thus, we find that finite-size effects at first-order
transitions can be effectively studied in terms of scaled cumulants.
These are easily extracted from data and have no size-dependence
within the context of the scaling theory. As a result, their study very
simply determines the limits of the theory, and tells us whether
the systems used in their extraction are large enough for infinite
volume physics to be extracted.

In the ten years that have passed since the first studies of
finite-size scaling at first-order phase transitions, the
computational power at the disposal of physicists has increased by
three orders of magnitude. The ability to convert this power into
precise knowledge of the physics contained in a Hamiltonian depends
upon  control of the systematics of finite-size effects.
This control is now complete.

\vfil\eject

\end{document}